\documentstyle[12pt,aasms4]{article}
\newcommand{\gam}{$\Gamma_{-25}$}
\newcommand{\ha}{H$\alpha$}
\newcommand{\hb}{H$\beta$}
\newcommand{\heh}{$\psi_{\rm He}$}
\newcommand{\htr}{H~II region}
\newcommand{\lam}{$\lambda$}
\newcommand{\mic}{$\mu$m}
\newcommand{\ngc}{NGC~891}
\newcommand{\nt}{[N~II]}

\newcommand{\oo}{[O~I]}
\newcommand{\ot}{[O~III]}
\newcommand{\otw}{[O~II]}
\newcommand{\pa}{Perseus Arm}
\newcommand{\st}{[S~II]}
\newcommand{\sth}{[S~III]}
\newcommand{\stnt}{\st/\nt}
\newcommand{\ts}{$T_*$}
\newcommand{\xm}{$X_{\rm edge}$(H$^0$)}
\newcommand{\z}{$|z|$}

\begin{document}

\title{The Warm Ionized Medium in the Milky Way and Other
Galaxies}

\author{John S. Mathis\footnote{Dept. of Astronomy, Univ. of Wisconsin;
475 N. Charter St., Madison WI 53706, USA;
mathis@uwast.astro.wisc.edu}}

\begin{abstract} There are now observations of several emission lines
from the ``Warm Ionized Medium" (WIM) (or, equivalently, the ``Diffuse
Ionized Gas") of the local ISM, the \pa\ in the Milky Way, and also in
several other galaxies. Interesting features of these observations
include the great strength of \nt\lam6563 ($\sim$ \ha\ in some cases)
and the fact that \st\lam6717/\nt\lam6583 is almost the same
($\sim$0.6 -- 0.7) in all locations and objects. Other line ratios
(e.g., \ot\lam5007/\hb) vary considerably.

This paper presents simple photoionization models that reproduce the
observed spectra, providing extra heating beyond that supplied by
photoionization is assumed (see Reynolds, Haffner, \& Tufte 1999). The
same extra heating was used for models of all stellar temperatures
being combined together, although it could easily depend on \ts.

With observed gas-phase abundances (not solar), the line ratios in the
local arm at $b$ = 0$\arcdeg$ are fitted with no extra heating and
(S/H) = 13 ppm, as opposed to solar ($\sim$ 20 ppm). Local gas
observed at $b$ = $-35\arcdeg$ requires extra heating of about \gam\ =
0.75, where \gam\ is the extra heating in units of $10^{-25}$ erg
H$^{-1}$ s$^{-1}$. In the \pa\, there are similar results: little
extra heating is required at \z\ = 500 pc, and \gam\ = 3.0 is needed
at \z\ = 1.2 kpc. To fit the observations, the gas-phase composition
in the \pa\ must be reduced as required by the Galactic abundance
gradient observed for \htr{}s. The requirements for \ngc\ (the best
observed other galaxy) at \z\ = 1 kpc and 2 kpc are similar to the
\pa: little or no extra heating near the plane (1 kpc in this case)
and \gam\ $\sim$ 3 at \z\ = 2 kpc. In \ngc\ there is also an increase
of \lam5007/\ha\ with \z\ that can only come about if most of the
ionizing radiation is supplied by very hot stars (type O4: \ts\ $\sim$
50000 K). Either their radiation must propagate from the plane to high
\z\ through very little intervening matter, or else the stars are
located at high \z. The total power requirement of the extra heating
is $\lesssim$ 15\% of the power to photoionize the WIM without extra
heating.

Extra heating enhances \otw\lam3727/\hb. Figure 1 shows that there is
a spread in the predicted values, but the ratio can serve as a useful
diagnostic of extra heating. The [S~III]\lam9065,9531 lines (see
Figure 2)are not useful in diagnosing extra heating.

\end{abstract}

\section{Introduction} It is well known that very diffuse ionized gas
is a major component of the interstellar medium (ISM) of the Milky Way
(Reynolds 1991, 1993). In some (but not all) other galaxies about 40\%
of the \ha\ is diffuse (Rand 1996; Hoopes, Walterbos, \& Greenawalt
1996; Domg\"orgen \& Dettmar 1997; Greenawalt et al. 1998; Otte
\& Dettmar 1999), as opposed to H~II regions (although it is not
completely clear that the definition of WIM is the same for
extragalactic objects as for the Galaxy.) The diffuse gas will be
referred to as the ``Warm Ionized Medium" (WIM) in this paper,
although it is also called ``the Reynolds layer" for the Milky Way gas
and also the ``diffuse ionized gas." The properties of the WIM,
especially regarding its heating, are reviewed by Reynolds, Haffner,
\& Tufte (2000).

The source(s) of the ionization of the WIM have been debated, but
photoionization by radiation from OB stars seems to be the only viable
source for the bulk of the material. Even for the local Milky Way, the
power required for the ionizations is very large: 10 -- 15\% of the
energy produced by OB stars, or all of the total power from
supernovae. It is not obvious how the ionizing radiation from OB stars
can penetrate the diffuse H~I that is observed in all directions from
the Sun. However, it seems likely that OB stars ionize enough of the
neutral material surrounding them to allow the required fraction of
ionizing radiation to escape (Miller \& Cox 1993; Dove, Shull, \&
Ferrera 2000). Since the Sun does not possess this ionizing power,
there are apparently no directions in our sky that allow ionizing
radiation produced by the Sun to escape to great heights from the
Galactic plane.

The hypothesis that the WIM is photoionized by hot massive stars is
supported by the agreement of its spectrum with models of very dilute
H~II regions (e.g. Domg\"orgen \& Mathis 1994; Greenawalt, Walterbos,
\& Braun 1997; Martin 1997; Wang, Heckman, \& Lehnert 1998; Sembach et
al. 2000). The main reason for the general agreement is that the
observed spectrum is heavily biased to low stages of ionization (\nt,
\st), with weak lines of the higher stages that are abundant in
ordinary H~II regions ([O~III], [S~III]). This bias is a natural
feature of nebulae photoionized by comparatively dilute radiation
fields. The models predict the observed anticorrelation of \nt/\ha\
and \st/\ha\ with the intensity of
\ha. However, [O~I]\lam6300 is weak in the only two Galactic WIM
regions for which it has been measured (Reynolds 1985, Reynolds et al
1998), both in the galactic plane. This line is strong only at the
outer edges of H~II regions, where there is both O$^0$ (with abundance
locked to H$^0$ by resonant charge exchange) and warm free
electrons. Present models of the local WIM (e.g., Sembach et al. 2000)
assume that some of the ionizing radiation escapes from the Galaxy, so
the resulting \lam6300 is not detected. The model reflects this
leakage by being matter bounded, with the fraction of neutral H,
$X$(H$^0$), $\sim$0.1 at the outer edge instead of $\sim$1.

The heating of the WIM is less well understood than the
ionization. Reynolds \& Cox (1992) suggested that some form of heating
in addition to that resulting from photoionization might be
required. At present there are several new observations that make the
case for extra heating very strong (Reynolds, Haffner, \& Tufte 1999),
and suggest that the extra heating, relative to the photoelectric
heating, decreases with density. Thus, photoionization heating
dominates in dense regions, such as near the midplane or in bright
H~II regions, but extra heating may become stronger than
photoionization heating in the very diffuse WIM.

The present paper will interpret several sets of observations in a
general way. The emphasis will be to compare observations relatively
close to the plane of the respective locations with those farther from
the plane. Photoionization models will be constructed with simple
geometries (i.e., volume averages of uniformly dense nebulae.)
Combining the photoionization models from assumed individual stars in
order to mimic the actual complicated situation will result in
conclusions that are plausibility arguments rather than detailed
models of the real situations. Firm conclusions can be reached as
regarding heating requirements beyond photoionization. More elaborate
modeling seems unjustified in view of the geometrical complexity, such
as fractal structure, that is found in the actual ISM.

The new observations that I have chosen to model are discussed in \S2
and listed in Table 1. The general characteristics of the
photoionization models are discussed in \S3, and results regarding
individual regions are in \S4. Discussion are given in \S5 and a
summary in \S6.

\section{Observations} %%%%%%%%%%

Older observations of the Milky Way WIM within the local spiral arm
have been summarized and modeled in Sembach et al. (2000). Their
models are close to those of Domg\"orgen \& Mathis (1994). This paper
will discuss the lines [S~II]\lam6717, hereafter ``[S~II]";
[N~II]\lam6583, hereafter ``[N~II]"; [O~III]\lam5007, hereafter
``\ot"; \oo\lam6300, hereafter ``\oo", and the quantity \heh\ $\equiv$
(He$^+$/He)/(H$^+$/H), the fractional ionization of He relative to H,
derived from He~I
\lam5876/H$\beta$. I will fit line ratios \nt/\ha, \nt/\st, \oo/\ha,
\ot/\ha, and the abundance ratio \heh\ in various objects. This paper
addresses the \st/\nt\ ratio, while previous papers (Domg\"orgen \&
Mathis 1994; Sembach et al. 2000) considered the absolute values of
the lines.

Excepting the local WIM, where there is good spatial resolution,
ratios from near galactic midplanes are not suitable for modeling
because of contamination by relatively dense H~II regions, including
radiation scattered by dust. The emission becomes too faint for
reliable measurements at large distances from the plane. 

New observations of the local WIM and for the Perseus Arm (Haffner,
Reynolds, \& Tufte 1999) were taken with the Wisconsin \ha\ Mapper
(WHAM) near the Galactic plane and also down to $b=-35\arcdeg$. The
\pa\ WIM can be identified by the velocity shift produced by galactic
rotation, and I will consider only \z\ = 500 pc and 1200 pc in the
\pa\ in order to avoid the relatively dense H~II regions near $b =
0\arcdeg$. Haffner et al. (1999) found \nt/\st\ $\sim$0.6 in the local
ISM at all $b$.

Reynolds et al. (1998) have measured \oo\ in the local Galactic
plane. In the Perseus Arm they observed it at two radial velocities
with strengths \oo/\ha\ = 0.044, and also $<$0.012, at \z\ = 330 pc. I
adopted \oo/\ha\ = 0.028 at \z\ = 500 pc as a reasonable
compromise. There are as yet no measurements of [O~III] for local gas
at $b=-35\arcdeg$ and for the \pa. 

While there are many observations of various galaxies, I choose to
concentrate on \ngc\ because it has the greatest variety of observed
lines of any galaxy. However, we must recognize that the observations
used here are from very deep spectra (Rand 1997, 1998) taken at one
slit position perpendicular to the plane at the location of an
especially bright filament. The physical conditions at the location at
a given \z{} were not typical of the entire galaxy. The average of the
line ratios at the same \z\ on the opposite sides of the galaxy was
used. The asymmetry above and below the plane is not large, so the
average is physically meaningful.

The adopted observations and other quantities explained below are
given in Table 1 for each of the regions that are modeled. The adopted
observed value is an average of a substantial number of individual
values near a particular value of $z$.

There are common properties of the spectra of the local WIM, the \pa,
and the WIM in \ngc\ that must be explained. One of the most striking
is the constancy of \stnt\ with \z. It is readily explained by a
constant (S/N) plus extra heating (Reynolds et al. 1999). However,
without extra heating a strong variation in the (S/H) with \z, highly
correlated with a different strong correlation of (N/H), is needed
(Rand 1998; Reynolds et al. 1999). The reason is that the \nt\ and
\st\ are not produced in precisely the same region: S$^+$ is much more
confined to the outer regions of the Str\"omgren sphere than is N$^+$.

There are substantial differences among the three regions. The local
WIM and, by assumption, the \pa\ have modest \oo\ and \heh\ (Reynolds
et al. 1998; Reynolds \& Tufte 1995), while \ngc\ has appreciable
\oo, larger \heh\ at high \z, and, especially, large measured
\ot/H$\beta$.

\section{Modeling the Observations} %%%%%%%%%

My approach was to compute individual photoionization model H~II
regions, each with a single central star and a specific value for the
``ionization parameter,'' $U$, which uniquely determines the
ionization structure of the model. The parameter $U$ is defined in
terms of a uniform density, $n_e$, by $U\equiv L/(4\pi R_S^2\,cn_e)$,
where $R_S$ is the Str\"omgren radius and $L$ is the H-ionizing photon
luminosity of the exciting star. I use a related quantity, $q\equiv
L_{50}f^2n_H$, where $f$ is the filling factor and $L_{50} \equiv
10^{-50}L$(photons s$^{-1}$). It is easy to show that $U=K\,q^{1/3}$,
with $K=[10^{50}\alpha_B^2/(36\pi (1+y)c^3)]^{1/3}$, where y is the
He/H ratio by number and $\alpha_B$ the recombination coefficient of H
to levels $n\ge2$. In order to vary $q$, I varied $f$ and assumed a
constant $L$ and $n$(H). The ionizing spectra were based on the
atmosphere models of Kurucz (1991). The nebular models were calculated
with the code used in Domg\"orgen \& Mathis, with atomic constants
updated to those in Pradhan \& Peng (1995).

It is important to realize that each model is characterized by the
{\em mean} ionization parameter, $q$, but contains within it a whole
range of variations of the {\em local} ratio of (photon/electron)
densities, since the stellar radiation field becomes weaker near outer
portions of the model. One cannot associate a particular measure of
$n_{\rm photon}/n_e$, such as might be inferred from observations of
the emission measure and distance from ionizing sources, with the
model that has that same average value of
$n_{\rm photon}/n_e$. However, volume-averaged models probably mimic
the complicated physical situation of the real ISM as well as any
assumed radial variation of density, or even a guess regarding the
three dimensional structure of the gas density. The actual ISM is
clumped on scales covering several orders of magnitude, and the
radiative transfer through it is very poorly represented by any
current models. A volume average seems to be a reasonable average. The
assumption of uniform density becomes serious for integrations along
particular lines of sight piercing the model.

Most models were assumed to have a spatially uniform mass density and
rate of extra heating. This case represents a constant deposition of
heating per H nucleus, independent of the state of ionization or other
physical conditions. Each volume-averaged model shows how uniform
extra heating per H nucleus affects the averaged spectrum. The results
cannot be used to determine the dependence of the extra heating rate
on $n_e$, as has been considered in the
\pa\ by Reynolds et al. (1999). Such a prediction would involve
detailed assumptions regarding the radial geometry of the nebula
around the exciting star(s).

The final fit to the observations was determined by combining two or
three individual H~II region models with possibly different values of
$q$. The agreement of the combination with the observations was made
by minimizing the statistic
\begin{equation}
\chi^2\equiv\sum_{i,k}\left\{\left[\frac{(Obs_i-a_k\,P_{k,i})}
{\epsilon_i}\right]^2 \right\}\:, \end{equation} 
\noindent where $i$
refers to the quantity to be fitted (such as \nt/\ha,) $k$ to
individual \htr{} models in the combination, and the $\epsilon_i$ are
discussed below. The $Obs_i$ are the observed ratios; $P_{k,i}$ is the
volume-averaged ratio in each model. The minimization was accomplished
by adjusting the $a_k$, the fraction of the ionizing photons
contributed by the individual model $k$. Only three individual models
were considered for any combination, but usually only two were
satisfactory. In view of the arbitrary nature of the geometry of the
individual \htr\ models (see below), including more individual models in
the combination would not have increased physical understanding.

In order to assess the quality of fit of models, I assign ``errors,"
$\epsilon_i$, to each quantity to be fitted. This ``error" is not an
assessment of the actual true uncertainty, but rather a fitting
parameter that reflects how closely I require the final superposition
of individual photoionization models to conform to the adopted
observational values. The $\epsilon_i$ are listed in Table 1 for each
ratio in each position.

Another approach to fitting the observations might be to estimate the
numbers of stars of various spectral types within clusters that are
exciting the WIM, to combine their luminosities, and then to compute
individual photoionized models with that stellar input. This procedure
is somewhat arbitrary as well. Sometimes the stars in a cluster act as
a collective source of radiation, as for a supershell (e.g., Dove et
al. 2000), but often stars produce their own individual \htr{}s that
contribute to the WIM. The ionizing radiation from very young clusters
is likely to be absorbed by the local gas, producing a rather bright
\htr, and contributing to the WIM through leakage of some of its
ionizing radiation. When the observed intensity is along a path
spanning an entire galaxy, the actual geometry is very complex.

Using rays piercing the outer regions of models might be just as
appropriate as volume averages, especially at large \z\ if the
exciting stars are found at smaller \z. This case represents looking
through the outer edges of the ionized region. I investigated using
this type of averaging, and it sometimes achieved excellent fits to
the observations. However, the results presented here are all volume
averages because the geometry of real nebulae seems so convoluted that
volume averages seem more appropriate. A major problem with using
integrations along paths is that the radial density distribution
within the model must be assumed arbitrarily. Mathis and Rosa (1991)
showed that volume averages of models with varying $q$ can well
represent the spectra of various regions within individual H~II
regions.

Other input parameters are the abundances of He, C, N, O, Ne, and S,
all relative to H. Values for each region are discussed in \S\ref{res}
below. The abundance of Ne has a minimal effect on the observed
ratios: a 10\% increase in Ne decreases \ot\ by 1.7\% and the other
ratios by $\le\ $1.2\%. Carbon has almost no effect on the results
because it is not a major coolant in the ionized regions, as opposed
to the neutral ISM, which lacks the powerful coolants [O~II], [N~II],
and recombination cooling. If (S/H) is increased, with the other
abundances held constant, \nt\ decreases slightly because of the extra
cooling. \st\ and \sth\ increase as (S/H)$^{0.8}$, resulting in
\st/\nt\ $\propto$ (S/H). For a similar reason, \st/\nt\ is almost
inversely proportional to (N/H).

The fraction of neutral H at the outer edge of the model, \xm, must
also be assumed. Radiation-bounded models have \xm\ = 0.9, at which
point the nebular temperature, if there is no extra heating, is
decreasing rapidly. Continuing to \xm\ = 0.95 gives almost identical
results, but is probably no closer to reality than \xm\ = 0.9 because
of the neglect of the dynamical effects that are present near
ionization fronts. I also considered gas-bounded models with \xm\ =
0.1, representing leakage of some ionizing radiation to larger \z. The
fraction of escaping ionizing photons depends on $q$, since low values
of $q$ imply a very gradual transition at the outer edge of the
model. For the range of $q$ I considered ($-2.6\ge\log(q)\ge-4.8$),
the fraction of escaping photons is $\sim$ 20 -- 55\%, respectively,
when \xm\ = 0.1.

The rate of extra heating per H nucleus, in units of 10$^{-25}$ erg
H$^{- 1}$ s$^{-1}$, is described by the quantity \gam.  The extra
heating is assumed to take place only out to $X$(H$^0$) = 0.5. The
same value of \gam\ was used for models of all stellar temperatures
being combined together, although \gam\ could easily depend on
\ts. The heating and power requirements are discussed under the
various regions whose spectra were fitted (\S\ref{res}). The power
required is considered in \S5.

The parameters that have major effects on the results at a given value
of $q$ are \ts, \xm, \gam, and, marginally, (O/H). The effects follow
intuition. Increasing \ts\ shifts the ionization balance to higher
ions. Even low values of $q$ show measurable \ot\ ($\sim$ 0.1 \hb) if
\ts\ = 50,000 K, while no models with \ts\ = 35,000 K produce
appreciable \ot. Decreasing \xm\ strongly decreases the emissions from
low stages of ionization, especially \oo\lam6300, which is strongly
concentrated to the outer edges of the models. For instance, models
with \xm\ = 0.1 have \lam6300 a factor of $\sim$5 weaker than those
with \xm\ = 0.9. Increasing (O/H) lowers the general level of
forbidden lines by providing extra cooling through \otw\lam3727.
Extra heating increases all emissions by increasing the nebular
temperature. However, \oo\lam6300/\st\lam6717 is almost independent of
extra heating because the line emissivities depend upon
temperature-dependent collision strengths and ionization fractions,
despite the somewhat different Boltzmann factors.

Dust is a source of uncertainty that does not affect the results
significantly. WIM models have much lower values of $q$ than bright
\htr{}s, implying a larger fraction of neutral H at a particular
fraction of $R_S$. Thus, in WIM models the opacity contributed by H is
relatively large. At the ionization edge of H, the absorption cross
section of an H atom is 6.3\,10$^{-18}$ cm$^2$ and of dust is $\sim$
3\,10$^{-21}$ cm$^{-2}$ H$^{-1}$ (e.g., Mathis, Rumpl, \& Nordsieck
1977). Thus, if the fraction of H$^0$ is $> 5\,10^{-3}$, as it is over
most of the volume for these low values of $q$, the opacity of H$^0$
is more than 10 times the dust opacity.

\section{Results\label{res}} %%%%%%%%

In this section we discuss the constraints placed on the physical
nature of the successful combinations for local gas, the \pa, and for
\ngc.

\subsection{The Local ISM} %%%%%%

For CNO abundances in the local ISM I ignored solar values and used
the observed {\em gas-phase} (O/H) = 319 per 10$^6$ H (Meyer, Jura, \&
Cardelli 1998); (N/H) = 75 ppm (Meyer, Cardelli, \& Sofia 1997); (C/H)
= 140 ppm (Sofia, Fitzpatrick, \& Meyer 1998). These give a larger
(N/O) than solar, presumably because an appreciable amount of O, but
little or no N, is incorporated in silicates and oxides in grains.
For the local ISM, the assumed abundances are very similar to those
called ``Warm Neutral Medium" in Sembach et al. (2000). The (S/H)
required to fit the WIM spectrum at $b=0\arcdeg$ is 13 ppm, so (S/N) =
0.17. Solar (S/N) = 0.22 $\pm$ 0.08 (Grevesse, Noels, \& Sauval 1996).
The local abundances at $b=-35\arcdeg$ were assumed to be the same as
in the plane because it seems very unlikely that significant amounts
of N are incorporated into dust grains (Whittet et al. 1998; Tielens
at al. 1999). Unless N and S can be released by destruction of grains
above the galactic plane, it seems unlikely that their gas phase
abundances will increase with \z.

For $b$ = 0$\arcdeg$, several combinations fitted without extra
heating. The best fit, matching the observations virtually perfectly,
was produced by 56\% of the ionizing photons from models with \ts\ =
35,000 K, (spectral type O9 V; Vacca, Garmany, \& Shull 1996), 12\%
from 40,000 K (spectral type O7 V) with \xm\ = 0.1, and the rest from
\ts\ = 45,000 K (type O5.5). The predictions of the combination are 
listed in Table 1. The relatively high stellar temperature required
probably represents the hardening of the stellar radiation that leaks
from denser H~II regions because of the absorption of the softer
photons. Other good combinations used as little as 32\% of the
ionizing photons from \ts\ = 35,000 K, with the other ionizing photons
divided between \ts\ = 40,000 K radiation bounded and gas bounded
models. The only problem with these models is their prediction that
\lam5007/\hb\ $\sim$ 0.08, while the observed is 0.18. Models with
\ts\ = 45,000 K fitted all of the adopted observations
almost perfectly. For any particular combination there is a range of
possible values of $q$.

Even a small amount of extra heating at $b$ = 0$\arcdeg$ is
problematic with the abundances adopted here. The major challenge for
the models mentioned above is to avoid overproducing \nt\ in the
plane, and extra heating makes the problem worse. Heating as low as
\gam\ = 0.4 increases \nt/\ha\ to 0.65, while the observations give
0.60 $\pm$ 0.02 (Haffner et al. 1999). Reducing the assumed (N/H) is
not an option because the ratio is observed (Meyer et al. 1997).

For the local arm ($v$(LSR) $\sim$0 km s$^{-1}$), the spectrum at $b$
= $-35\arcdeg$ differs from that at $b$ = 0$\arcdeg$ by having
\nt/\ha\ $\sim$ 0.80 in place of 0.60 at $b$ = 0$\arcdeg$ and \st/\ha\
increased by about the same ratio. There are no measurements of \oo\
or \ot. Extra heating is required to produce the stronger \nt/\ha\ and
\st/\ha. A combination of models similar to those for $b$ =
0$\arcdeg$, but with \gam\ = 0.75, fits the observations: 50\% of the
ionizing photons from \ts\ = 35,000 K, and the remainder from 45,000
K. The predictions of the combination are listed in Table 1.  If we
assume that the \oo, \ot, and \heh\ are the same at $b = - 35\arcdeg$
as for 0$\arcdeg$, we can eliminate models with \gam\ = 0. They
produce either too little \nt\ or else much too much He~I, with the
best fit being an uncomfortable accommodation of the two errors,
predicting
\heh\ $\sim$0.63. The upper limit in the plane is \heh\ $\le$
0.4. Extra heating at the level of \gam\ = 0.7 can produce an
excellent fit to all ratios, including those assumed to be similar to
the plane (\oo, \heh, and \ot). Observations of \heh, \oo, and \ot\
would be very useful.

We have seen that a combination of \ts\ = 35,000 K and hotter stars
produces good fits to the spectrum of the local gas. Domg\"orgen \&
Mathis (1994) used a single \ts\ = 38,000 K; Sembach et al. (2000)
used 35,000 K.

\subsection{The Perseus Arm} %%%%%%%%%

Abundances in the \pa\ were estimated by fitting the
observations of the WIM rather than using the galactic abundance
gradient. They were (C, N, O, Ne, S)/H = (85, 45, 250, 71, 8)
ppm, respectively. If they had been assumed to be the local gas phase
abundances, reduced by a factor proportional to $D_G^{0.066}$ ($D_G$
is the distance from the Galactic center), they would have been (85,
50, 213, 85, and 9.3) ppm, respectively\footnote{The exponent 0.066 is
the mean of those found in Afflerbach, Churchwell, \& Werner (1997)
for O, N, and S, very similar to the gradient in B star abundances
found by Gummersbach et al. (1998). The distance to the \pa\ from the
Sun was assumed to be 2.5 kpc and the observations to the \pa\ were at
$l$ = 145$\arcdeg$. The reduction factor in this case is 0.72.}. In
view of the assumptions within the models, the differences between the
two sets of abundances are not significant.

Like the results for the local spiral arm, several combinations of
models fit the observations in the \pa\ at \z\ = 500 pc. No extra
heating is required. Examples of successful combinations are 55\% of
ionizing photons from \ts\ = 35,000 K and the rest from 40,000 K. The
predictions (see Table 1) are excellent. Another very good fit arises
from 35,000 K (70\%) combined with 45,000 K, \xm\ = 0.1. The best fit
is with 66\% of the ionizing photons from models with \ts\ = 35,000 K,
4\% from 50,000 K, \xm\ = 0.9, and 30\% from 50,000 K, \xm\ =
0.1. However, such very hot stars do not seem appropriate for the
\pa. The range of reasonably successful combinations illustrates the
difficulty of being very specific about the ionizing spectrum.

At \z\ = 1200 pc, there are no observations of either \oo\ or \ot, but
\nt/\ha\ $\sim$1 and \st/\ha\ = 0.6. Models without extra heating fail
to produce enough \nt; even a \ts\ = 50,000 K model produces \nt/\ha\
of only 0.75. Increasing (N/O) and (S/O) to provide the required \nt\
and \st\ is not attractive because the abundances would have to
greater than solar, while the (O/H) would need to kept low in order to
keep the nebula hot.

Models with \ts\ = 35,000 K (50\% of photons) combined with \ts\ =
45,000 K fits the observations at \z\ = 1200 pc well (see Table 1)
with extra heating of \gam\ = 3. The models produce relatively large
amounts of \nt\ and \st, the opposite of those at 500 pc. Models with
lower densities in the outer regions also produce an excellent fit.

The power requirements of the extra heating will be discussed in
\S\ref{disc}.
\subsection{NGC 891} %%%%%%%%
The major difference between \ngc\ and the Galaxy is that in \ngc,
\lam5007/\hb\ increases with \z\ to
$\sim$0.75 at \z\ = 2 kpc, as opposed to
$\le$0.2 in the plane of the Milky Way. Other
differences are the relative large strength of \oo/\ha\ ($\sim$0.1
in \ngc\ vs. $<$ 0.03) and \heh\ (0.55 in \ngc, vs. 0.4 in the \pa).

In order to avoid the emission from dense H~II regions near the plane,
the smallest \z\ I considered in \ngc\ is 1 kpc. The \pa\ at that \z\
requires extra heating. However, \ngc\ has about the same \nt/\ha\ at
1 kpc as in the plane of the Milky Way, so I assumed no extra heating
in \ngc\ at \z\ = 1 kpc. Even low values of \gam\ ($\sim$ 0.4) make it
difficult to obtain enough \oo\ without overproducing \nt\ and \st.

The abundances in \ngc\ were assumed to be (C, N, O, Ne, S)/H = (85,
45, 250, 71, 8) ppm, respectively. The \pa\ is the same except it used
(O/H) = 213 ppm. The N, O, and S were found by fitting the spectra at
both \z\ = 1 and 2 kpc, with no extra heating at \z\ = 1 kpc. The C
and Ne have little influence on the spectra and were assumed to be the
same relative to N as in the Galaxy.

It is not possible for local gas phase abundances, (N, O)/H = (75,
320) ppm, to fit the rather low \nt/\ha\ in \ngc\ at the same time as
the \heh\ and \ot. With this high an (N/H), \nt\ is overproduced by
the hot stars needed for \ot. Assuming appreciable extra heating in
\ngc\ at \z\ = 1 kpc makes all of these problems worse. Such
extra heating cannot be ruled out if the compositions are taken to be
free parameters. We would require \ngc\ to have (N/H) $<$ 45 ppm
(about half solar) when integrated across the galaxy, and a gas phase
(N/O) $\le$ 70\% of the local value.

Good fits for \ngc, \z\ = 1 kpc are found with $\sim$50\% of the
photons from \ts\ = 35,000 K and the rest from \ts\ = 45,000 or 50,000
K (Types 05.5 or earlier). The cool stars are needed to reduce the
\heh\ to $\sim$0.5, and the hot stars in order to produce enough \nt\
and \heh. Using \ts\ = 40,000 K (type O7.5) for the hot stars does not
fit because its \ot\ is too weak. If stars with \ts\ $\ge$ 45,000 K
are included, the major problem in \ngc\ is not in fitting the \nt, as
it was for the \pa, but in producing enough \oo. The best combination
is a composite of three models: \ts\ = 35,000 K (48\%), \ts\ = 50,000
K, \xm\ = 0.9 (32\%), and \ts\ = 50,000 K, \xm\ = 0.1 (20\%). It
produced an excellent fit ($\chi^2/\nu=0.2$, where $\nu$ is the number
of line ratios to be fitted = 5.) Table 1 lists its predictions.
However, the range of combinations with (say) $\chi^2/\nu\le0.6$ is
large. There are successful combinations with values of $q$ ranging
over an order of magnitude, all with the \ts\ $\ge$ 45,000 K for the
hot stars.

At \z\ = 2 kpc for \ngc\ we assume the same abundances as for \z\ = 1 kpc,
which is an important constraint because the spectra are rather
different: \nt\ and \st\ $\sim$ \ha\ instead of 0.6 \ha\ at 1 kpc, \ot\ =
0.75 \hb\ as opposed to 0.55, and \heh\ = 0.85 vs 0.55.

Because of \heh, $\sim$90\% of the ionizing radiation must arise from
very hot stars (\ts\ $\sim$45,000--50,000 K) instead of $\sim$ 50\% at
1 kpc. To produce high \heh, higher values of \ts\ are required in the
WIM than in H~II regions because the H and He have much large
fractions of neutrals at any particular fraction of the nebular
radius. These ions compete for O$^+$ ionizing photons ($h\nu\ge35.5$
eV) much more successfully than in the more compact nebulae.  In
addition, strong extra heating is required at 2 kpc in \ngc.  Extra
heating rate \gam\ = 3 is the least that provided a good fit to the
spectrum; it required 14\% from \ts\ = 35,000 K and the remainder from
\ts\ = 50,000 K (73\% radiation bounded; 13\% from gas bounded.) The
hot stellar temperature could be reduced if more extra heating (\gam\
= 4) is allowed. With
\gam\ = 4.5, \ts\ = 45,000 K can be used in place of 50,000 K.

The presence of very hot stars at large \z\ in \ngc\ is not
unreasonable. There is a great deal of dust found well out of the
plane (Howk \& Savage 1997), and in other edge-on spirals as well
(Howk \& Savage 1999). If there is dust, there can be star formation
and associated hot stars found nearby.

Must these very hot stars be at large \z, if the WIM at large \z\ is
photoionized by starlight? In order to simulate ``chimneys'' with hot
stars near the plane, I computed models with central voids of
0.6$R_S$. At a given \nt/\ha, the volume-averaged
\ot/\hb\ is about the same or only slightly weaker than in the case of
uniform density. However, these models depend on having no gas between
the exciting stars and the outer regions. For low values of $q$, H$^0$
and He$^0$ efficiently absorb photons with $h\nu\ge$ 35.5 eV, the
ionization edge of O$^{+2}$. Unless there is almost a void in the gas
below 2 kpc and exciting stars near the plane, there would be so
little O$^{+2}$ that extra heating will increase \nt\ and \st, but
\ot\ would be too weak.

\subsection{Predictions of [O~II] and [S~III] Lines\label{otwo}}

The strongest line from the WIM is almost surely \otw\lam3727 because
of the observed weakness of \ot, the abundance of oxygen, and the
efficiency of \otw\ as a coolant. Figure 1 summarizes the predictions
of \otw\lam3727/\hb\ of the models of various stellar temperatures.
The solid line is for \ts\ = 50,000 K; the short dashes are for 45,000
K; the dotted is for 40,000 K; the long dashes are for 37,500 K; the
long/short dashes are for 35,000 K. At various points along each line,
error bars indicate the range of line strengths among the various
values of $q$ considered for each model. The lowest value of
\lam3727/\hb\ corresponds to the lowest log($q$) (= $-4.8$), but the
highest occurs at various values around log($q)\sim-3.5$.

The composition of the models shown by lines and error bars were (C,
N, O, Ne, and S) = (85, 45, 250, 71, and 8) ppm, respectively, which
is the composition used for \ngc.

The points in Figure 1 show the mean \lam3727/\hb\ for \ts\ = 50,000 K
(circles), 40,000 K (squares), and 35,000 K (triangles), for nebulae
containing more heavy elements than those used for the lines in the
figure: (C, N, O, Ne, and S) = (141, 75, 320,118, 13) ppm,
respectively. These are the abundances adopted in this paper for the
local WIM.  The difference in the points and lines shows that \lam3727
is influenced by the cooling from other ions, while the rough
agreement between the two compositions shows that \lam3727 is a
major coolant, especially if there are large amounts of extra
heating. Our composite models for the WIM used roughly
equal amounts of \ts\ = 35,000 K and 50,000 K, which is not far from
\ts\ = 40,000 K as regards \lam3727/\hb.

From Figure 1 we see that there is a general correlation of \lam3727
with heating, but with a large scatter, especially at low extra
heating rates. The \lam3727/\hb\ cannot be used to estimate low
heating rates accurately, but it can distinguish between \gam\ = 0 and
\gam\ = 3. From observations of \lam3727/\hb, Figure 1 can be used to
estimate the mean amount of extra heating in the WIM. The figure
cannot be used to determine the density dependence of the extra
heating, since the assumed geometry of the models does not correspond
to the unknown density distribution and clumpy structure of the real
WIM.

Figure 2 is similar to Figure 1, except that it shows
[S~III]\lam(9065+9531)/\ha\ for the same models as in Figure 1. The
values are almost proportional to (S/H) for large rates of extra
heating. We see that [S~III] is not a good diagnostic of physical
conditions within the WIM.

\section{Discussion \label{disc}}

In spite of significant differences among the spectra of the WIM in
the local ISM, the \pa, and \ngc, there a considerable similarity in
the simple photoionization models fitting them. No extra heating is
required, and little tolerated, at low \z, where \nt/\ha\ = 0.5 --
0.6. At larger \z, where \nt/\ha\ $\sim$1, \gam\ $\sim$ 3 is needed in
at least the outer regions of the plasma where N and S are largely
singly ionized. There is little evidence from these photoionization
models regarding the relative amounts of extra heating in the regions
near the exciting star(s), where the photoionization heating dominates
the extra heating if it is assumed to be uniform.

At large \z\ in the \pa\ and \ngc, very hot stars (\ts\ $\ge$ 45,000
K) are needed to explain the comparatively large \nt/\ha, \st/\ha, and
\oo/\ha\ ratios observed. These high stellar temperatures might
represent the hardening of the radiation escaping from denser \htr{}s
closer to the plane, since the softer photons are absorbed first by
the H and He. In \ngc very hot stars, and not hardening
of the stellar radiation, are needed to explain the \ot\ observed at
\z\ = 2 kpc. For the low values of $q$ required for the WIM, 
the gas producing the hardening would absorb the O$^+$ ionizing
photons required for the \lam5007 line. These hot stars could be in
the plane of the galaxy, powering a superbubble (Dove et al. 2000) if
there is very little gas between them and \z\ = 2 kpc.

These combinations of models fail to predict precisely what values of
the ionization parameter, or $q$, are required to produce the observed
spectra. For any particular object there is a rather wide range of
values that are satisfactory. The other geometrical uncertainties,
such as using volume integrals, simplistic stellar atmospheres, or
requiring the same heating for all stellar temperatures at a given
location, make predictions of the dilution of the radiation field
quite speculative.

The propagation of the ionizing photons from the hot stars to the WIM
may be the most serious objection to the idea the photoionization is
the dominant process operating in the WIM, since only a very small
amount of neutral H is completely opaque. The fractal nature of the
ISM is shown in molecular clouds (e.g., Elmegreen \& Falgarone 1996;
Stutzki et al. 1998) and in H~I (Green 1993). The percolation of
ionizing radiation through hierarchical density structures has not yet
been investigated quantitatively at an adequate level, but probably
ionizing photons from an O star can penetrate large distances through
the lowest density regions within the fractal ISM.

While photoionization from stars seems to be very plausible for
providing most of the ionization of the WIM, it is also very likely
that other sources contribute. Candidates are shocks (Raymond 1992)
caused by supernovae, as well as turbulent mixing layers (Slavin,
Shull, \& Begelman 1993) produced by shear instabilities of outflowing
very hot gas. The spectra of both of these phenomena are really
produced by photoionization, but the photons are emitted from hot
plasma instead of stellar photospheres. The contributions of 
metagalactic radiation upon the WIM at the largest distances from
the plane may be important, especially for producing \ot. While
\ot\ is surprisingly bright relative to \ha\ at large \z\ in \ngc, the
total power required for it is not large. The energy
requirements of the WIM at modest \z\ ,say, $\sim$ 1 kpc, are the most
problematic constraint on the fraction of the WIM that is contributed
by these alternative processes.

The energy requirements of the extra heating can be determined from
the details of the models. The detailed models show that the
photoelectric heating from a \ts\ = 50,000 K star is 30\% of the power
required for the ionization. This number is approximately $kT_*/
\chi(\rm H)$. This is expected, since $kT_*$ is the mean energy of a
black body spectrum above the ionization threshold $\chi(\rm H)$. For
\ts\ = 35,000 K, the extra heating power at \z\ = 1200 pc in the \pa\
(i.e., \gam\ = 3) is 65\% of the photoelectric heating, integrated
over the volume, and 30\% of the photoelectric heating for \ts\ =
50,000 K. The power requirement of the extra heating is more modest in
comparison to the total power needed, because we must then include the
ionization energy in addition to the heating. At \z\ = 1200 pc the
extra heating is 10\% total power requirement of the WIM. The extra
heating power, integrated over $z$ from 500 pc to 1200 pc, is $\sim$
5\% of the total power requirement of the WIM. The power requirements
for still larger \z\ are even smaller, since the intensity of the WIM
decreases rapidly with height.

While \lam3727 is the largest single coolant when extra heating is
small, it is only about 20\% of the total cooling, with \nt\ and \sth\
not far behind it. The \lam3727 is, of course, very sensitive to extra
heating because of its high excitation energy as compared to the other
optical lines.  Its importance relative to the other forbidden lines
increases with \gam\ because its emissivity increases faster than that
of the red lines.  However, its overall share of the cooling remains
about constant because of the onset of cooling by H$^0$ from
collisional excitation of the Lyman lines. The collisional cooling
exceeds \lam3727 at about \gam\ = 1.5, but the precise boundary is set
by the abundances and stellar temperatures.

It is well known that [C~II] 158 \mic\ is a major coolant in the
neutral ISM, as shown by models of photo-dissociation regions (PDRs)
on the outer parts of molecular clouds (e.g., Kaufman et al. 1999).
The {\em COBE} satellite showed (Wright et al. 1991) that 158 \mic\ has
a strength greater than the sum of all other infrared lines. The
present models show that [C~II] 158 \mic\ is not a major coolant in
the WIM, although most of the gas phase C is singly ionized. The
reason is that other coolants become dominant, while 158 \mic\ is not
boosted much by the relatively high temperature in the WIM as compared
to PDRs. A simple calculation shows the effect. Let $j_{158}(T)$ be
the emissivity (erg cm$^{-3}$ s$^{-1}$) of 158 \mic\ and similarly for
\nt\lam6583. We find a mean $T$ from setting $I_{\lambda 6583} /
I_{{\rm H}\alpha} = j_{\lambda 6583}(T) / j_{{\rm H}\alpha}(T)$, where
the $j$'s involve the appropriate ionic abundances. The most favorable
case for 158 \mic\ is using the largest (N/H), 75 ppm, and the lowest
$I_{\lambda6583}/ I_{{\rm H}\alpha}$, 0.60, resulting in $T$ = 7555
K. In this case, $j_{158}/j_{6583} = 0.33$, which may seem
significant. However,
\lam6583 provides only a small part of the total cooling, according
to models that predict \lam6583/\ha\ = 0.6. For these
models, the cooling of \lam6583 is $\sim$12\% of the total, so 158
\mic\ is only 4\% of the total. The \pa\ and \ngc\ models have
lower (N/H) and (O/H) and, therefore, have hotter nebulae, making the 158
\mic\ even less important. 

The nebular temperature is raised by both extra heating and a shift of
the contribution of hot stars to the ionization, so extra heating
cannot be clearly identified until it dominates the photoionization
heating.  Figure 1 shows that this takes place for \otw\lam3727 at
\gam $\sim$ 2. Below this value, the influences of stellar
temperatures and abundances of coolants are comparable with those of
extra heating. However, He$^+$/H$^+$ is unaffected by extra
heating. Very hot stars are needed if \heh\ approaches unity,
especially for the low-density situation in the WIM. \ngc\ is an
excellent example of this situation.

It is significant that the combinations of models at low \z\ in the
Milky Way cannot tolerate much extra heating, while extra heating is
required at large \z. 

The heating might arise from turbulence (Minter \& Spangler 1997;
Minter \& Balzer 1997). For models with uniform density, this extra
heating can also be assumed to represent heating that is uniform per
unit volume, such as magnetic reconnection (Raymond 1992;
Gon\c{c}alves et al. 1993; Birk, Lesch, \& Neukirch 1998; Vishniac \&
Lazarian 1999). See Haffner et al. (1999) for discussion and
references.

Heating by PAHs is produced by a change of their state of ionization
(Bakes \& Tielens 1994). Variations in the interstellar radiation
field seen along the line of sight should affect the spectra of the
near-infrared ``Unidentified Infrared Bands" (UIBs) in the diffuse ISM
and also the extra heating if PAHs are responsible for
it. Observations of the spatial variations of the UIBs, and
correlations between these changes and the extra heating as shown by
\nt/\ha\ and \st/\ha, will be crucial in determining if PAHs are the
main source of extra heating.  Complete uniformity of the UIB spectra,
suggesting that the distribution of stages of ionization of PAHs is
uniform in various environments, would not support heating by PAHs.

While there are many free parameters in the combinations of models,
there are also strong constraints. The composition was taken to be the
same at high and low \z, and subject to observational values and the
galactic abundance gradients for the local gas and the \pa. Up to 5
line ratios were fitted with 2 stellar temperatures and one relative
fraction of ionizing photons. There are some combinations of
observations that cannot be predicted by photoionization models.  Our
range of $q$ considered ($-4.80\le\log(q) \le -2.60$) implies
\nt$\ge$ 0.4\ha, with the minimum occurring for cool stars, \gam\ = 0,
and gas bounded models with large central voids. My models all have
\oo\ $\le$ 0.6\hb, with the maximum occurring for hot stars, the
largest \gam\ I considered (4.5), the lowest $q$, and a large central
void. We never predict an \ot\lam5007/\hb\ $<$ 0.15 combined with
\heh\ $\ge$0.9 (i.e., \ts\ $\sim$50,000 K because of the generally low
stage of ionization present, or small $q$).

At the outer edge of the model, forbidden-line cooling (mainly from
[S\,II]) cannot keep the temperature below $\sim$ 10000 K if there is
strong extra heating that is constant per H nucleus. Collisional
ionization of H begins to produce a low but important fraction ($>$
1\%) of H$^+$, along with corresponding electrons. In this case, large
amounts of [S~II]\lam6717 and [O~I]\lam6300 are produced ($\sim$ twice
\ha) without any photoionization at all. Since such strong \oo\ has
not been observed, we conclude that collisional ionization and
collisional excitation induced by the extra heating cannot produce
most of the observed \ha. Clearly, a physical theory is needed to
determine the level of H ionization at which various processes provide
extra heating, and to estimate the conditions under which it becomes
negligible in comparison to photoelectric heating. The present models
arbitrarily end the extra heating when the fraction of H$^0$ is $>$
0.5, but values of 0.1 or 0.9 give essentially the same
results. However, the model cannot be continued to completely neutral
H, which never occurs with constant extra heating because of
collisional ionization of H.

\section{Summary}

As has been shown in past work, photoionization models can explain the
observed spectra of the WIM. In this paper similar nebular models were
applied to the WIM spectra from several rather different regions: the
local spiral arm, the \pa\ at two values of \z, and \ngc\ along a well
observed strip perpendicular to the galactic plane.

The combinations of nebular models assume that only two different
stellar atmospheres provide the ionizing radiation. These stellar
atmospheres (Kurucz 1991) are valid only in broad outline because they
do not treat the stellar winds and non-LTE effects, and do not have
the composition of the nebular gas. I regard the numerical values of
the stellar temperature associated with each model with considerable
caution because of these failings. However, WIM spectra basically
count ionizing photons of various energies, so a distribution of stars
with various temperatures can be simulated with two discrete
values. Another question is whether the WIM is best characterized by
combining stars of various temperatures and then computing the nebular
spectrum, rather than combining nebular spectra from various
individual stars. Other assumptions involve the geometry used in
interpreting the model spectra, such as whether to use volume averages
of integrals along various lines of sight through the model
nebulae. Radiative transfer modeling of the fractal structure of the
ISM is required before one can be confident about the propagation of
the ionizing radiation from the source stars near the plane to large
\z.

An important modification to the simplest picture of photoionizing the
WIM is that extra heating beyond that supplied by photoionization is
required (Reynolds et al. 1999). The diagnostics of this requirement
are the large \nt/\ha\ ratio that is observed at large \z\ in both the
\pa\ and, especially, in \ngc. Another feature is the observed
constancy of \st/\nt at $\sim$0.6. There are several possible sources
of this extra heating. It must be relatively unimportant for low
\z. An increase with density less rapid than $n^{2}$ helps fulfill this
condition.

Excepting \ngc\ at large \z\ (2 kpc in this paper), the models that
are successful in fitting the WIM require both hot stars ($T_* \sim$
45,000 -- 50,000 K) and cool stars ($T_* \sim$ 35,000 K), each
providing comparable numbers of the ionizing photons. The cool stars
are needed for the observed rather low values of \nt\/\ha\ and
\st/\ha.  The hot stars are needed for \ot\ and He$^+$.

\ngc\ at \z\ = 2 kpc is very interesting in that it shows very strong
\ot\ and He$^+$, far above the plane.  The combinations of models show
that very hot stars (\ts\ $\sim$ 50,000 K) are needed, most plausibly
at high \z. If the exciting stars are found near the plane, their
radiation must propagate through almost a vacuum up to the \z\ = 2 kpc
region.

The (N/H) required for the local gas is about what is observed from UV
absorption studies, and that required for the \pa\ is about what is
expected on the basis of local gas phase measurements plus galactic
abundance gradients as observed in both stars and in H~II regions.
The fact that photoionization requires reasonable assumptions
regarding the exciting spectra and compositions of the WIM makes it a
very plausible candidate for explaining the principal physics of the
WIM.

With \nt\ strengthened by extra heating, the \st\ follows if the (N/S)
ratio is the same at high \z\ as near the plane. This constancy is
expected if neither element is strongly bound to grains in dilute
gas. The (S/N) required for the \pa\ is about 80\% of solar.

The abundances in both the \pa\ and the region sampled in \ngc\ seem
to be lower in N, S, and O than the observed gas-phase abundances in
the local arm. The accuracy of the composition predictions from these
combinations of simple models is compromised by the model assumptions,
but the result is so clear that it is almost surely real.

This work has benefited from conversations with many colleagues,
especially Ron Reynolds. Blair Savage, Bob Benjamin, Matt Haffner, and
several colleagues at the workshop on the WIM held in Green Bank, WV,
23/24 September 1999 provided stimulating ideas and criticisms. I
liked the suggestions from the anonymous referee.

%%%%%%%%%%%%%%%%%%%%TABLES%%%%%%%%%
\pagestyle{empty}
\begin{deluxetable}{cccccc}
\tablenum{1}
\tablecolumns{6}
\tablecaption{Observations and Predictions}
\tablehead{
\colhead{}
&\colhead{$\frac{\textstyle{\rm [N II]}\lambda6563}
 {\textstyle\rm H\alpha}$}
&\colhead{$\frac{\textstyle{\rm [S II]}\lambda6717}
 {\textstyle{\rm [N II]}\lambda6563}$}
&\colhead{$\frac{\textstyle{\rm [O I]}\lambda6300}
 {\textstyle\rm H\alpha}$}
&\colhead{$\psi_{\rm He}$\tablenotemark{a}}
&\colhead{$\frac{\textstyle{\rm [O III]}\lambda5007}
 {\textstyle\rm H\beta}$}
 } %end of \tablehead

\startdata
\multicolumn{6}{c}{Local ISM, $b=0\arcdeg$}\\
Observed &0.60&0.60&0.025&0.38&0.18\\
Adopted $\epsilon$\tablenotemark{b}&0.017&0.1&0.017&0.08&0.08\\
Predicted\tablenotemark{c}&0.60&0.61&0.024&0.48&0.21\\
\hline
\multicolumn{6}{c}{Local ISM, $b = -35\arcdeg$}\\
Observed &0.80&0.65&$\cdots$&0.38&$\cdots$\\
Adopted $\epsilon$&0.017&0.10&$\cdots$&0.08&$\cdots$\\
Predicted\tablenotemark{c}&0.81&0.66&0.044&0.38&$0.11$\\
\hline
\multicolumn{6}{c}{\pa, \z\ = 500 pc}\\
Observed &0.50&0.60&0.028&0.40&$\cdots$\\
Adopted $\epsilon$&0.025&0.06&0.017&0.07&$\cdots$\\
Predicted\tablenotemark{c}&0.49&0.61&0.023&0.40&0.10\\
\hline
\multicolumn{6}{c}{\pa, \z\ = 1200 pc}\\
Observed &1.00&0.60&$\cdots$&0.55&$\cdots$\\
Adopted $\epsilon$&0.025&0.06&$\cdots$&0.07&$\cdots$\\
Predicted&0.97&0.61&0.080&0.56&$\cdots$\\
\hline

\tablebreak

\multicolumn{6}{c}{\ngc, \z\ = 1 kpc }\\
Observed &0.58&0.63&0.08&0.55&0.32\\
Adopted $\epsilon$&0.025&0.06&0.017&0.07&0.07\\
Predicted&0.58&0.66&0.062&0.58&0.30\\
\hline
\multicolumn{6}{c}{\ngc, \z\ = 2 kpc}\\
Observed &1.0&0.63&0.10&0.85&0.75\\
Adopted $\epsilon$&0.017&0.06&0.01&0.05&0.07\\
Predicted&1.01&0.60&0.10&0.88&0.67\\
\enddata
\tablenotetext{a}{$\psi_{\rm He}\equiv$ (He$^+$/He)/(H$^+$/H).}\\
\tablenotetext{b}{$\epsilon$ is the one-$\sigma$ accuracy with
which the models are expected to reproduce the observations;
see equation (1).}
\tablenotetext{c}{The prediction is with a combination of model
H~II regions with properties given in \S\ref{res} in the text.}
\end{deluxetable}

\newpage

\begin{figure}
\plotone{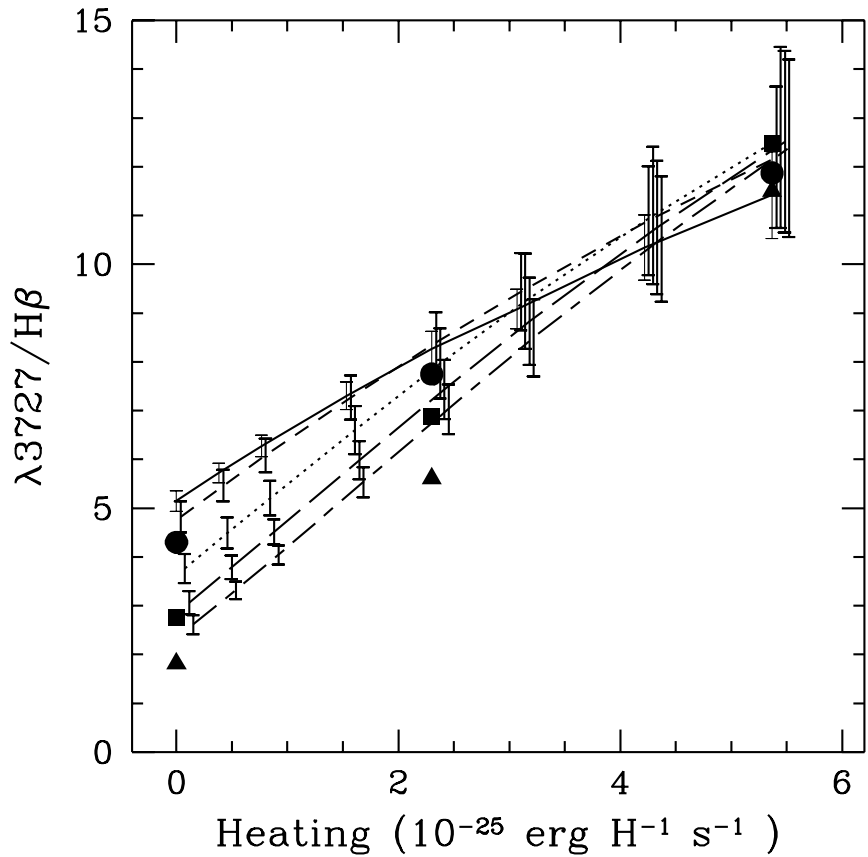}
\caption{
\setlength{\baselineskip}{0.6cm}
 The strength of [O~II]\lam3727/\hb, plotted against the rate
of extra heating in the models, for models with two different
compositions and and various temperatures of exciting stars. The lines
are all for the composition (C, N, O, Ne, S)/H = (85, 45, 250, 71, 8)
ppM. The solid line shows \ts\ = 50,000 
K; the dotted, 40,000 K; the long dashed, 37,500 K; the
long-short dashed, 35,000 K. The ``error bars'' indicate the maxima
and minima for the models at various values of the excitation
parameter, which determines the amount of H$^0$ at a particular
fraction of the Str\"{o}mgren radius. Those for cooler stars are
displaced slightly to the right for clarity.
The points show the same as the lines, except that the composition is
(C, N, O, Ne, S)/H = (141, 75, 320,118, 13) ppM. The circles are for
50,000 K; the squares for 40,000 K; the triangles for 35,000 K. We see
that extra heating strongly increases the strength of \lam3727. The
composition, through changing the cooling, also enters the line ratio.
}
\end{figure}

\newpage
\begin{figure}
\plotone{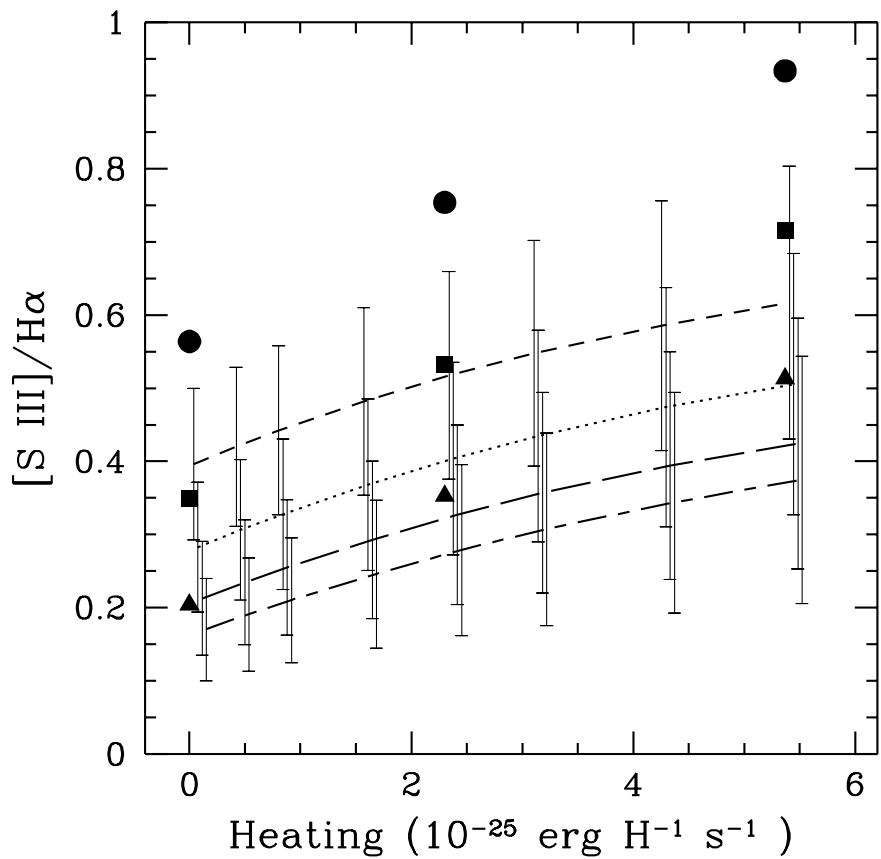}
\caption{
The same as Figure 1, except for the ratio
[S~III]\lam(9065+9531)/\ha. We see that this line ratio is not a 
diagnostic of extra heating.
}
\end{figure}

\end{document}